\documentclass[12pt]{article}

\usepackage{amsfonts,amssymb,bm,cite,amsmath}
\usepackage[dvipdfmx]{graphicx}
\usepackage{array, booktabs}
\usepackage{braket}
\usepackage[title]{appendix}

\newcommand {\beq}{\begin{equation}}
\newcommand {\eeq}{\end{equation}}
\newcommand {\beqa}{\begin{eqnarray}}
\newcommand {\eeqa}{\end{eqnarray}}
\newcommand {\n}{\nonumber \\}
\newcommand {\tr}{\mbox{tr}}
\newcommand {\Tr}{\mbox{Tr}}

\newcommand {\del}{\partial}

\renewcommand{\theequation}{\thesection.\arabic{equation}}

\begin{document}

\setlength{\oddsidemargin}{0cm}
\setlength{\baselineskip}{7mm}

\begin{titlepage}
\renewcommand{\thefootnote}{\fnsymbol{footnote}}
\begin{normalsize}
\begin{flushright}
\begin{tabular}{l}
February 2020
\end{tabular}
\end{flushright}
\end{normalsize}

~~\\

\vspace*{0cm}
    \begin{Large}
       \begin{center}
         {Information geometry encoded in bulk geometry}
       \end{center}
    \end{Large}
\vspace{1cm}

\begin{center}
           Asato T{\sc suchiya}\footnote
            {
e-mail address :
tsuchiya.asato@shizuoka.ac.jp}
           {\sc and}
           Kazushi Y{\sc amashiro}\footnote
           {
e-mail address : yamashiro.kazushi.17@shizuoka.ac.jp}\\
      \vspace{1.5cm}

{\it Department of Physics, Shizuoka University}\\
                {\it 836 Ohya, Suruga-ku, Shizuoka 422-8529, Japan}\\
         \vspace{0.5cm}
 {\it Graduate School of Science and Technology, Shizuoka University}\\
               {\it 3-5-1 Johoku, Naka-ku, Hamamatsu 432-8011, Japan}

\end{center}

\vspace{3cm}

\begin{abstract}
\noindent
We study how information geometry is described by bulk geometry in the gauge/gravity correspondence. We consider a quantum information metric that measures the distance between the ground states of a CFT and a theory obtained by perturbing the CFT.
We find a universal formula that represents the quantum information metric
in terms of back reaction to the AdS bulk geometry.
\end{abstract}
\vfill
\end{titlepage}
\vfil\eject

\setcounter{footnote}{0}

\section{Introduction}
\setcounter{equation}{0}
\renewcommand{\thefootnote}{\arabic{footnote}}

Emergence of space-time (geometry) is considered to play an essential role
in constructing quantum theory of gravity.  Indeed, it is observed in various contexts
including the AdS/CFT correspondence or the gauge/gravity correspondence\cite{Maldacena:1997re,Gubser:1998bc,Witten:1998qj},
where the bulk direction on the gravity side emerges as the scale of renormalization group
on the field theory side \cite{Gubser:1998bc,Witten:1998qj,Susskind:1998dq,deBoer:1999tgo}.
This observation motivates one to reconstruct full bulk geometry from field theory.
The Ryu-Takayanagi formula \cite{Ryu:2006bv} gives a hint to this problem.
It relates entanglement entropy of a region in space on which a field theory is defined
to
the area of a minimal surface in the bulk whose boundary agrees with that of the region.
Thus, it gives a relationship between quantum information theory and bulk geometry.

In this paper, to further gain insights into this problem, we consider
information metric in quantum information theory other than entanglement entropy,
and investigate how they are encoded in bulk geometry.
We represent information metric in terms of
back reaction to the AdS bulk geometry, which is determined by dynamics of gravity.
The geometrical quantity associated with the information metric
is local in the bulk direction, while the minimal surface
associated with entanglement entropy is not.
Information metrics have been studied in the context of the AdS/CFT correspondence
in \cite{MIyaji:2015mia,Bak:2015jxd,Trivella:2016brw,Chen:2018vkw,Karar:2019wjb,Nozaki:2012zj,
Lashkari:2015hha,Aoki:2017bru,May:2018tir}.

The authors of \cite{MIyaji:2015mia} considered a CFT and a theory that is obtained
by perturbing the CFT by an operator and calculate an information metric that measures
the distance between the ground states
of these two theories. They examined a gravity dual of a filed theory that is obtained by gluing
the above two theories and found that the information metric is represented by the volume of a hypersurface
that is the time slice in the bulk which ends on the time slice on the boundary.
Further developments in this direction have been made in \cite{Bak:2015jxd,Trivella:2016brw,Chen:2018vkw,Karar:2019wjb}. A different type of information metric has been investigated in
\cite{Lashkari:2015hha,Aoki:2017bru,May:2018tir}.

We consider the same set of two filed theories and the same information metric
as those in \cite{MIyaji:2015mia}.
Then, we examine two gravity duals, one of which is dual to the CFT and the other of which is dual to
the perturbed CFT.
The bulk geometry in the latter gravity dual gains the
back reaction caused by the perturbation.
We find a formula (\ref{formula 2}) that expresses the information metric by
deviation of the volume of a hypersurface in the bulk from
that in the case of the AdS geometry.
This formula is universal in the sense that it holds for each case in which
the perturbation is given by a scalar, vector
or tensor operator. The formula is new findings in this paper.

This paper is organized as follows.
In section 2, we briefly review information metrics in field theories.
In section 3, we consider a CFT and a theory that is obtained by perturbing the CFT by
a scalar primary operator. We calculate the information metric that measures
the distance between the ground states of these two theories, and associate it
to the on-shell action on the gravity side by using the GKP-Witten relation.
In section 4, we evaluate the back reaction to the AdS geometry caused by the perturbation
and find the formula that relates the information metric to
deviation of the volume of the hypersurface in the bulk from
that in the case of the AdS geometry.
In sections 5 and 6, we examine the cases in which CFTs are perturbed by vector and tensor operators, respectively. We find that the same formula holds as in the scalar case.
Section 7 is devoted to conclusion and discussion. The Ricci tensor and the scalar curvature are calculated in appendix.

\section{Information metric in field theory}
\setcounter{equation}{0}

We consider a field theory defined on $\mathbb{R}^d$, whose coordinates are
$(\tau, \vec{x})$, where $\tau$ is the Euclidean time and $\vec{x}$ are the $(d-1)$-dimensional
space  coordinates. The wave function of the ground state $|\Omega\rangle$
in the theory is represented
by a path integral over a time interval from $-\infty$ to $0$ as follows:
\begin{align}
\langle \tilde{\psi} | \Omega \rangle
=\frac{1}{Z^{1/2}} \int_{\psi(0,\vec{x})=\tilde{\psi}(\vec{x})}
 {\cal D}\psi
\exp\left[-\int_{-\infty}^{0} d\tau \int d^{d-1}x \
{\cal L} \right]  \ ,
\label{wave fn of ground state}
\end{align}
where the value of the field $\Psi$ is fixed to $\tilde{\psi}(\vec{x})$ at $\tau=0$, and
$Z$ is the partition function of the theory. Note that the ground state is normalized:
\begin{align}
\langle \Omega | \Omega \rangle = 1 \ .
\end{align}

We further consider two theories, the theory 1 and the theory 2,  with the same field content defined by the lagrangians
$\mathcal{L}_1$ and $\mathcal{L}_2$, respectively.
We denote the ground states of these two theories
by $|\Omega_1\rangle$ and $|\Omega_2\rangle$, respectively.
Then, by glueing the wave functions for these ground states given in (\ref{wave fn of
ground state}), we can represent the inner product between the ground states
in terms of a path integral
as
\begin{align}
\langle \Omega_2 | \Omega_1 \rangle
=\frac{1}{(Z_1Z_2)^{1/2}} \int {\cal D}\psi
\exp \left[-\int d^{d-1}x \left(\int_{-\infty}^0d\tau \
{\cal L}_1
+\int_0^{\infty}d\tau \ {\cal L}_2 \right) \right] \ ,
\label{inner product}
\end{align}
where $Z_1$ and $Z_2$ are the partition functions of the theory 1 and the theory 2,
respectively.

We denote the difference of the two lagrangians by $\delta\mathcal{L}$.
Namely, we have
\begin{align}
{\cal L}_2={\cal L}_1+\delta {\cal L} \ .
\end{align}
We rewrite the inner product (\ref{inner product}) in terms of $\delta\mathcal{L}$
as
\begin{align}
\langle \Omega_2 | \Omega_1 \rangle
=\frac{\left\langle \exp \left[-\int_0^{\infty}d\tau
\int d^{d-1}x \ \delta  {\cal L}\right]
\right\rangle_1}
{\left\langle \exp \left[-\int_{-\infty}^{\infty}d\tau
\int d^{d-1}x  \ \delta {\cal L}\right]
\right\rangle^{1/2}_1}  \ ,
\label{inner product in terms of deltaL}
\end{align}
where $\langle \; \rangle_1$ stands for the vacuum expectation value taken with respect to
the theory 1:
\begin{align}
\langle {\cal O} \rangle_1
=\frac{1}{Z_1} \int {\cal D}\psi \ {\cal O} \
e^{-S_1}
=\langle \Omega_1 | {\cal O} | \Omega_1 \rangle   \ .
\label{}
\end{align}
We expand (\ref{inner product in terms of deltaL}) in terms of $\delta\mathcal{L}$ up to
$\mathcal{O}(\delta\mathcal{L}^2)$ as \cite{MIyaji:2015mia}
\begin{align}
\langle \Omega_2 | \Omega_1\rangle
= & 1-\frac{1}{2}\int_{0}^{\infty}d\tau\int_{-\infty}^{0}d\tau'
\int d^{d-1}x \int d^{d-1}x' \
\langle \delta{\cal L}(\tau,\vec{x})\delta{\cal L}(\tau',\vec{x}')\rangle_1 \ ,
\label{perturbative expansion}
\end{align}
where we have assumed
\begin{align}
\langle \delta\mathcal{L}(\tau,\vec{x}) \rangle=0
\label{1pt function}
\end{align}
and
the time reversal symmetry
\begin{align}
\langle \delta\mathcal{L}(\tau,\vec{x})\delta\mathcal{L}(\tau',\vec{x})\rangle
=\langle \delta\mathcal{L}(-\tau,\vec{x})\delta\mathcal{L}(-\tau',\vec{x})\rangle \ .
\label{time reversal symmetry}
\end{align}

We further assume that $\delta\mathcal{L}$ takes the form
\begin{align}
\delta\mathcal{L}=\phi(\vec{x})\mathcal{O}(\tau,\vec{x}) \ ,
\label{deltaL}
\end{align}
where $\phi(\vec{x})$ is a source independent of $\tau$ and $\mathcal{O}(\tau,\vec{x})$
is a local operator. Then,  using (\ref{perturbative expansion}) and (\ref{deltaL}),
we introduce the information metric $\mathcal{G}$ which is known as Fisher's metric
and measures the distance between the ground states of the two theories:
\begin{align}
\mathcal{G}=\frac{1}{T}(1-\langle \Omega_2 | \Omega_1\rangle)
=\int d^{d-1}x \int d^{d-1}x' \
\mathcal{G}_{\vec{x},\vec{x}'} \ \phi(\vec{x})\phi(\vec{x}')
\label{information metric}
\end{align}
with
\begin{align}
\mathcal{G}_{\vec{x},\vec{x}'}
=\frac{1}{2T}\int_{0}^{\infty}d\tau\int_{-\infty}^{0}d\tau'
\langle {\cal O}(\tau,\vec{x})
{\cal O}(\tau',\vec{x}')\rangle_1 \ ,
\label{information metric components}
\end{align}
where $T$ is the volume of time direction.

\section{Information metric as on-shell action}
\setcounter{equation}{0}

Let us consider a case in which the theory 1 is a CFT and
$\mathcal{O}(\tau,\vec{x})$ in (\ref{deltaL}) is a scalar primary operator
with the conformal dimension $\Delta$ in the CFT.
Namely, the theory 2 is obtained by
perturbing the CFT  by the scalar primary operator.
In what follows, the quantities in the theory 1 are labeled `CFT' instead of `1', while those in the theory 2
have no labels. For instance,
\begin{align}
\mathcal{L}=\mathcal{L}_{CFT}+\phi(\vec{x})\mathcal{O}(\tau,\vec{x})
\label{lagrangian with perturbation}
\end{align}
and so on.

The one-point function of the primary operator $\mathcal{O}$ vanishes,
which implies that (\ref{1pt function}) is satisfied.
The two-point function of the primary operator $\mathcal{O}$ takes the form
\begin{align}
\langle {\cal O}(\tau,\vec{x}){\cal O}(\tau',\vec{x}')\rangle_{CFT}
=\frac{C_{\Delta}}{(\epsilon^2+(\tau-\tau')^2
+(\vec{x}-\vec{x}')^2)^{\Delta}}  \ ,
\label{2pt correlation function}
\end{align}
where $C_{\Delta}$ is a normalization constant and a UV cutoff $\epsilon$ has been
introduced. We see from (\ref{2pt correlation function}) that (\ref{time reversal symmetry})
is satisfied.

The information metric (\ref{information metric}) and (\ref{information metric components})
reads
\begin{align}
\mathcal{G}
& = \frac{1}{T}(1- \langle \Omega | \Omega_{CFT}\rangle)\nonumber\\
&=\frac{1}{8}\int_{-\infty}^{\infty} ds
\int d^{d-1}x \int d^{d-1}x'
\frac{C_{\Delta}\phi(\vec{x})\phi(\vec{x}')}
{(\epsilon^2+s^2+(\vec{x}-\vec{x}')^2)^{\Delta}} \ .
\label{information metric 2}
\end{align}

Suppose that the CFT has a gravity dual defined on $AdS_{d+1.}$.
Throughout this paper,
we consider a situation in which the classical approximation is valid on the gravity side.
Because the information metric takes the form of the generating functional for
the two-point functions, it can be represented by the on-shell action for
the bulk field $\Phi$
dual to $\mathcal{O}$.

Here we introduce the following notations:
$
x^{\mu}=(x^0,x^i)=(\tau,\vec{x})$,
where $\mu=0,\ldots,d-1$ and $i=1,\ldots,d-1$,
and
$z^M=(z,x^{\mu})$.
We use the metric of $AdS_{d+1}$ in the Poincare coordinates, which takes  the form
\begin{align}
ds^2=G_{MN}dz^Mdz^N=\frac{1}{z^2}(dz^2+dx^{\mu}dx^{\mu}) \ .
\label{AdS metric}
\end{align}
We define a boundary hypersurface in $AdS_{d+1}$
by $z=\epsilon$, where $\epsilon$ was introduced in
(\ref{2pt correlation function}) as a UV cutoff. The CFT is viewed as defined on the boundary.

The action for $\Phi$ on the gravity side is
\begin{align}
S_M
= \frac{1}{2} \int d^{d+1} x \sqrt{G} \left( G^{MN} \del_M \Phi \del_N \Phi + m^2 \Phi^2 \right) \ ,
\label{S_M}
\end{align}
where we have presented only the quadratic terms in $\Phi$, which are needed in the following,
and
\begin{align}
m^2=\Delta(\Delta-d) \ .
\end{align}

The equation of motion for $\Phi$ is derived from (\ref{S_M}) as
\begin{align}
-\frac{1}{\sqrt{G}}\partial_{M}(\sqrt{G}G^{MN}
\partial_{N}\Phi) +m^2\Phi=0 \ .
\label{equation of motion for Phi}
\end{align}
The boundary condition for $\Phi$ is given by
\begin{align}
\Phi(z=\epsilon,\tau,\vec{x})=\epsilon^{d-\Delta}\phi(\vec{x}) \ .
\label{boundary condition for Phi}
\end{align}
The solution to (\ref{equation of motion for Phi}) satisfying the boundary condition
(\ref{boundary condition for Phi}) \cite{Witten:1998qj} is
\begin{align}
\Phi(z,x) &= \int d^{d}x' K (z, x-x') \phi(\vec{x}')  \ ,
\label{solution}
\end{align}
where $K$ is the so-called boundary to bulk propagator:
\begin{align}
K(z,x)=\frac{\alpha_{\Delta}z^{\Delta}}{(z^2+x^2)^{\Delta}}  \; \; \mbox{with} \;\;
\alpha_{\Delta}=\frac{\Gamma(\Delta)}{\pi^{\frac{d}{2}}\Gamma(\Delta-\frac{d}{2})} \ .
\end{align}
Note that $\Phi(z, x)$ is independent of $\tau$.

By substituting (\ref{solution}) into (\ref{S_M}), we evaluate the on-shell action for $\Phi$ as
follows:
\begin{align}
S_{on-shell}
&= \frac{1}{2} \int d^{d+1} x \del_M\left(\sqrt{G}  G^{MN}  \Phi \del_N \Phi \right)
\nonumber\\
&\;\;\; - \frac{1}{2} \int d^{d+1}x \sqrt{G}\Phi \left\{\frac{1}{\sqrt{G}} \del_M \left( \sqrt{G} G^{MN}  \del_N \Phi \right) - m^2 \Phi \right\}  \nonumber\\
&=
-\frac{1}{2} \int_{z=\epsilon} d^d x \ \epsilon^{-d+1} \ \Phi \del_z \Phi  \nonumber\\
& =
-\Delta \alpha_{\Delta}  \int_0^{\infty} d\tau \int_{-\infty}^{\infty}  ds \int d^{d-1} x  d^{d-1} x' \frac{\phi(\vec{x}) \phi(\vec{x}')}{(\epsilon^2
+ s^2+(\vec{x}-\vec{x}')^{2} )^{\Delta}}  \ ,
\label{on-shell action}
\end{align}
where we have used (\ref{equation of motion for Phi}) to obtain the third equality.
This would be the generating functional of $\phi(\vec{x})$ for two-point functions of
$\mathcal{O}$ so that we obtain
\begin{align}
C_{\Delta} = \Delta \alpha_{\Delta} \ .
\end{align}
Then, by comparing (\ref{on-shell action}) with (\ref{information metric 2}), we find
\begin{align}
S_{on-shell} = -4T\mathcal{G} \ .
\label{on-shell action and information metric}
\end{align}

\section{Back reaction to the AdS geometry}
\setcounter{equation}{0}
The theory (\ref{lagrangian with perturbation})
obtained by perturbing the CFT by the primary operator
would have a gravity dual where the geometry has
a back reaction to the AdS geometry, namely deviates from the AdS geometry.
We evaluate the back reaction up to $\mathcal{O}(\phi^2)$ in the following.

We parametrize the metric with the back reaction as
\begin{align}
ds^2=G_{MN}dz^Mdz^N=\frac{1}{z^2}(dz^2+g_{\mu\nu}(z,x)dx^{\mu}dx^{\nu})  \ ,
\label{metric}
\end{align}
with
\begin{align}
g_{\mu\nu}(z,x) = \delta_{\mu\nu} + h_{\mu\nu}(z,x) \ ,
\label{expansion of metric}
\end{align}
where $h_{\mu\nu}$ represent the back reaction to the AdS geometry and
start with $\mathcal{O}(\phi^2)$ contribution, which we will focus on.

The gravity action on the gravity side is given by
\begin{align}
S_{G} = \frac{1}{16\pi G_N} \left[\int d^{d+1} x \sqrt{G} \left(- R[G] + 2\Lambda \right)
- \int_{z=\epsilon} d^d x \sqrt{\gamma} (2K+\lambda) \right] \ ,
\label{gravity action}
\end{align}
where the cosmological constant $\Lambda$ is $\Lambda = \frac{-d(d-1)}{2}$,
the boundary cosmological constant $\lambda$ is $\lambda=-2(d-1)$, $\gamma$ is the induced metric on the boundary, and $K$ is the trace
of the extrinsic curvature. The scalar curvature $R[G]$ is defined in (\ref{scalar curvature}).

The Einstein equation is derived from (\ref{gravity action}) and (\ref{S_M}) as
\begin{align}
  R[G]_{MN} + d G_{MN} = 8\pi G_N \mathcal{T}_{MN}  \ ,
  \label{Einstein equation}
\end{align}
where the Ricci tensor $R[G]_{MN}$ are defined in (\ref{Ricci tensor}), and
$\mathcal{T}_{MN}$ are defined as
\begin{align}
\mathcal{T}_{MN} \equiv T_{MN} - \frac{1}{d-1} G_{MN} G^{IJ}T_{IJ}
\end{align}
with
\begin{align}
T_{MN}
  = \del_M \Phi \del_N \Phi - \frac{1}{2} G_{MN} \left( \del_L \Phi \del^L \Phi +  m^2 \Phi^2 \right) \ .
\end{align}

Each component of (\ref{Einstein equation}) takes the following form:
\begin{align}
& \Tr g^{-1} g'' - \frac{1}{z} \Tr g^{-1}g' - \frac{1}{2} \Tr g^{-1}g'g^{-1}g'
 = -16 \pi G_N \mathcal{T}_{zz}
 \ ,  \label{zz component} \\
& \nabla_{\mu} \Tr g^{-1} g' - \nabla^{\lambda} g'_{\lambda \mu} = -16 \pi G_N \mathcal{T}_{z\mu}  
\ ,   \label{zmu component} \\
&  g''_{\mu\nu} - g'_{\mu \lambda} g^{\lambda \sigma} g'_{\sigma \nu} + \frac{1}{2} \Tr \left(g^{-1} g' \right) g'_{\mu\nu} - (d-1) \frac{1}{z} g'_{\mu\nu} - \frac{1}{z} \Tr \left( g^{-1} g' \right) g_{\mu\nu} - 2 {\rm{Ric}}_{\mu\nu}(g) \n
  & = -16 \pi G_N \mathcal{T}_{\mu\nu}
\label{munu component}
\end{align}
with
\begin{align}
\mathcal{T}_{zz} &=\del_z \Phi \del_z \Phi + \frac{m^2}{ d-1} \frac{1}{z^2} \Phi^2
\ , \nonumber\\
\mathcal{T}_{z\mu} &= \del_z \Phi \del_{\mu} \Phi
\ , \nonumber\\
\mathcal{T}_{\mu\nu} &=  \del_{\mu} \Phi \del_{\nu} \Phi + \frac{m^2}{ d-1} \frac{1}{z^2} g_{\mu \nu} \Phi^2
\ ,
\end{align}
where the prime stands for the derivative with respect to $z$,
Tr is defined by $\mbox{Tr}A= \delta^{\mu\nu}A_{\mu\nu}$,
$\nabla_{\mu}$ is the covariant derivative with respect to the metric $g_{\mu\nu}$,
and ${\rm{Ric}}_{\mu\nu}[g]$ are the Ricci tensor for $g_{\mu\nu}$.

We expand
the left hand sides up to $\mathcal{O}(h_{\mu\nu})$ in order to evaluate
$h_{\mu\nu}$ up to $\mathcal{O}(\phi^2)$, since the righthand sides
of (\ref{zz component}), (\ref{zmu component}) and
(\ref{munu component}) are $\mathcal{O}(\phi^2)$.
Here, $h_{\mu\nu}$ is independent of $\tau$, because so is $\Phi$, and
we ignore total derivative terms with respect to $\vec{x}$, which will be justified shortly.
Then,
(\ref{zz component}) reduces to
\begin{align}
\mbox{Tr} h'' - \frac{1}{z} \Tr h' = -16 \pi G_N \left( \del_z \Phi \del_z \Phi + \frac{m^2}{ d-1} \frac{1}{z^2} \Phi^2 \right)  \ ,
  \label{eq: zz}
\end{align}
while (\ref{munu component}) reduces to
\begin{align}
h_{\mu\nu}'' - (d-1) \frac{1}{z} h_{\mu\nu}' - \frac{1}{z} \Tr  h' \delta_{\mu\nu}
  = -16 \pi G_N \left( \del_{\mu} \Phi \del_{\nu} \Phi + \frac{m^2}{ d-1} \frac{1}{z^2} \delta_{\mu \nu} \Phi^2 \right) \ .
  \label{eq: munu}
\end{align}
Taking the trace of (\ref{eq: munu}) yields
\begin{align}
\mbox{Tr} h'' - (2d-1) \frac{1}{z} \Tr h'
  = -16 \pi G_N \left( \del_{\mu} \Phi \del_{\mu} \Phi + \frac{m^2}{ d-1} \frac{d}{z^2} \Phi^2 \right)  \ .
  \label{eq: Trmunu}
\end{align}
The $00$ component of (\ref{eq: munu})  reads
\begin{align}
h_{00}''- (d-1) \frac{1}{z} h_{00}' - \frac{1}{z} \Tr  h'
  = -16 \pi G_N \frac{m^2}{ d-1} \frac{1}{z^2} \Phi^2 \ .
  \label{eq: 00}
\end{align}

By taking a linear combination of (\ref{eq: zz}), (\ref{eq: Trmunu}) and (\ref{eq: 00}),
we obtain
\begin{align}
  \tr h'' - \frac{d - 1}{z} \tr h' = -8 \pi G_N \left\{ \del_z (\Phi \del_z \Phi) - \frac{d-1}{z} \Phi \del_z \Phi \right\} \ ,
\label{trh''}
\end{align}
where $\mbox{tr}A=A_{ii}=\mbox{Tr}A-A_{00}$, and we have used (\ref{equation of motion for Phi}) and again ignored the total derivative terms with respect to $\vec{x}$.
Integrating (\ref{trh''}) leads to
\begin{align}
  \tr h' = -8 \pi G_N \Phi \del_z \Phi  \ ,
  \label{eq: trh(1)' scalar}
\end{align}
where the boundary condition $\lim_{z\rightarrow\infty}h_{\mu\nu}=0$ has been used.

By using the third equality in (\ref{on-shell action}) and (\ref{eq: trh(1)' scalar}),
we obtain
\begin{align}
  S_{on-shell} = \frac{1}{16 \pi G_N}\int_{z=\epsilon} d^d x \epsilon^{-d+1} \tr h'  \ .
  \label{eq: phi on-shell}
\end{align}
Here ignoring the total derivative terms in deriving (\ref{eq: trh(1)' scalar}) is justified.

By comparing (\ref{on-shell action and information metric}) and (\ref{eq: phi on-shell}),
we find a formula
\begin{align}
  \mathcal{G} = -\frac{1}{64 \pi G_N} \int_{z=\epsilon} d^{d-1} x\epsilon^{-d+1} \tr h'  \ .
\label{formula}
\end{align}
This formula represents the information metric in field theory in terms of
buck reaction to the AdS bulk geometry.
The righthand side of (\ref{formula}) is interpreted geometrically as follows.

\begin{figure}[t]
\centering
\includegraphics[width=10cm]{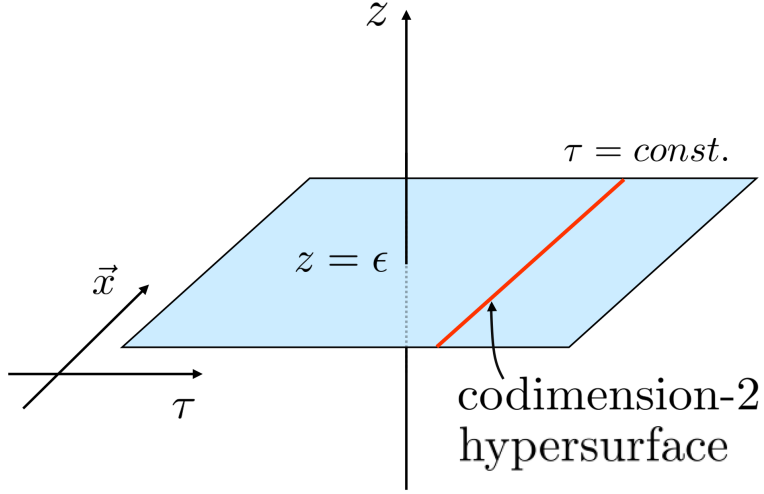}
\caption{A red line represents a codimension-2 hypersurface.}
\label{ccdimension-2}
\end{figure}

We consider a codimension-2 hypersurface specified by $z=\epsilon$ and $\tau=\mbox{const.}$
(see Fig.\ref{ccdimension-2}).
The induced metric on the hypersurface in the static gauge is given by
\begin{align}
    \gamma_{ij} &= \frac{\partial z^M}{\partial x^i} \frac{\partial z^N}{\partial x^j} G_{MN}\n
    & = \frac{1}{z^2} g_{ij}  \ .
\end{align}
The volume of the hypersurface is evaluated up to $\mathcal{O}(h_{ij})$ as
\begin{align}
  v &= \int_{z=\epsilon} d^{d-1} x \sqrt{\gamma}\n
  &= \int_{z=\epsilon} d^{d-1} x z^{-d+1} \sqrt{{\rm{det}}g_{ij}}\n
  &= \int_{z=\epsilon} d^{d-1} x z^{-d+1} \left(1 + \frac{1}{2}\tr h \right) \ .
\end{align}
We subtract the volume of the hypersurface in the AdS case where $h_{ij}=0$ and
denote the difference by $\delta v$:
\begin{align}
\delta v = \frac{1}{2}\int_{z=\epsilon} d^{d-1} x z^{-d+1} \tr h   \ .
\label{delta v}
\end{align}
By taking the derivative with respect to $z$, we obtain
\begin{align}
  \delta v' &= \frac{1}{2}\int_{z=\epsilon}
  d^{d-1} x \left( (-d+1)z^{-d} \tr h + z^{-d+1} \tr h' \right) \ .
\label{derivative of delta v}
\end{align}
While the first term in (\ref{derivative of delta v}) represents the canonical scaling of the volume,
the second term represents a nontrivial scaling of the volume and is proportional to
the righthand side of (\ref{formula}).
Thus, (\ref{formula}) is rewritten as
\begin{align}
\mathcal{G}=-\frac{1}{32\pi G_N} \delta v'_{nontrivial} \
\label{formula 2}
\end{align}
with
\begin{align}
\delta v'_{nontrivial} = \frac{1}{2}\int_{z=\epsilon}d^{d-1} x \
  z^{-d+1} \tr h'  \ .
\label{delta v nontrivial}
\end{align}

\section{Vector field}
\setcounter{equation}{0}
In this section, we extend the analysis in sections 3 and 4 to the case
of vector field.

We perturb a CFT by a U(1) vector current $\mathcal{J}^{\mu}(x)$: the counterpart of
the Lagrangian (\ref{lagrangian with perturbation}) is given by
\begin{align}
  \mathcal{L} = \mathcal{L}_{CFT} + a_{\mu}(\vec{x}) \mathcal{J}^{\mu}(x) \ ,
\end{align}
where the source $a_{\mu}(\vec{x})$ is
independent of the time, and $a_{0} = 0$. The 2-point function of $\mathcal{J}^{\mu}(x)$ is given by
\begin{align}
  \braket{\mathcal{J}_{\mu}(\tau,\vec{x})\mathcal{J}_{\nu}(\tau',\vec{x}')}= \frac{C_V}{(\epsilon^2 + |x-x'|^2)^{d-1}} J_{\mu\nu}(\epsilon,\tau-\tau',\vec{x}-\vec{x}') \ ,
\label{2pt function of current}
\end{align}
 where
 \begin{align}
  J_{\mu\nu}(\epsilon,\tau-\tau',\vec{x}-\vec{x}')=\delta_{\mu\nu} - \frac{2(x-x')_{\mu} (x-x')_{\nu}}{\epsilon^2 + |x-x'|^2}
  \ ,
\end{align}
and $C_V$ is a normalization constant. Note that (\ref{1pt function}) and (\ref{time reversal symmetry}) are satisfied.

The information metric, which is the counterpart of (\ref{information metric 2}), is
\begin{align}
  \mathcal{G}
  &=  \frac{C_V}{8} \int_{\infty}^{\infty}  ds \int d^{d-1} x \int d^{d-1} x' a^{i}(\vec{x})a^{j}(\vec{x}') \frac{J_{ij}(\epsilon,s,\vec{x}-\vec{x}')}{(\epsilon^2 +s^2+ (\vec{x}-\vec{x}')^2)^{d-1}} \ .
  \label{eq: information metric of vector}
\end{align}

A bulk field corresponding to $\mathcal{J}_{\mu}$ is a $U(1)$ gauge field $A_M$. The bulk
action $S_A$ for $A_M$ is given by
\begin{align}
  S_{A} = \frac{1}{4} \int d^{d+1} x \sqrt{G} F^{MN} F_{MN} \ ,
  \label{eq: action of vector}
\end{align}
where $F_{MN}=\del_M A_N - \del_N A_M$.
The equations of motion for $A_M$ derived from (\ref{eq: action of vector})
are
\begin{align}
  \frac{1}{\sqrt{G}} \del_M (\sqrt{G} F^{MN}) = 0 \ .
\label{EOM for vector}
\end{align}
Solving these equations around the AdS metric (\ref{AdS metric}) with a gauge
$A_z=0$ and a boundary condition $A_{\mu}(\epsilon,x)=a_{\mu}(\vec{x})$ leads
to \cite{Witten:1998qj}
\begin{align}
  A_{\mu} (z, x) &=
  \alpha_{V}\int d^d x' \frac{z^{d-2}}{(z^2 + |x-x'|^2)^{d-1}}J_{\mu}^{\;i}(z,\tau-\tau',\vec{x}-\vec{x}') a_{i}(\vec{x}')
  \label{eq: bb vector}
\end{align}
with
\begin{align}
\alpha_V = \frac{1}{2}\frac{\Gamma(d)}{\pi^{\frac{d}{2}} \Gamma(\frac{d}{2})} \ .
\end{align}
Note that $A_{\mu}(z, x)$ is independent of $\tau$ and that $A_{0} = 0$.

By substituting (\ref{eq: bb vector}) into (\ref{eq: action of vector}), the on-shell action
is obtained as
\begin{align}
  S_{A,on-shell} &= \frac{1}{2} \int d^{d+1}x \del_M \left(\sqrt{G} G^{KM} G^{LN} A_N F_{KL} \right)
  - \frac{1}{2} \int d^{d+1} x A_N \del_M (\sqrt{G} F^{MN} )\n
  &=  -\frac{1}{2} \int_{z=\epsilon} d^{d}x z^{-d+3} A^i F_{z i}  \nonumber\\
  & = -\frac{(d-2)\alpha_V}{2} \int^{\infty}_{-\infty} ds \int d^{d-1} x d^{d-1} x' a_{i}(\vec{x}) \frac{J_{ij}(\epsilon,s,\vec{x}-\vec{x}')}{(\epsilon^2 +s^2+ |\vec{x}-\vec{x}'|^2)^{d-1}} a_{j}(\vec{x}') \ .
  \label{on-shell action for vector}
  \end{align}
This on-shell action is the generating functional for the 2-point function
for (\ref{2pt function of current}) so that $C_V$ is determined as
$C_{V} = (d-2) \alpha_V$.
By comparing (\ref{eq: information metric of vector}) and (\ref{on-shell action for vector}),
we again obtain (\ref{on-shell action and information metric}).

We consider the back reaction to the AdS geometry. The bulk action consists of
the gravity part (\ref{gravity action}) and the gauge field part  (\ref{eq: action of vector}).
The Einstein equations derived from the bulk action are
(\ref{zz component}), (\ref{zmu component}) and (\ref{munu component}) with
\begin{align}
 \mathcal{T}_{zz} & =\frac{d-2}{d-1} z^2 g^{\mu\nu} F_{z\mu} F_{z\nu} - \frac{1}{2(d-1)} z^2  F_{\mu \nu} F_{\mu\nu} \ , \nonumber\\
 \mathcal{T}_{z\mu} & = z^2 g^{\alpha \beta} F_{z\alpha} F_{\mu \beta} \ , \nonumber\\
 \mathcal{T}_{\mu\nu} & =  z^2 \left( F_{z \mu} F_{z \nu} - \frac{1}{d-1} g_{\mu\nu} g^{\alpha \beta} F_{z \alpha} F_{z\beta} + g^{\alpha \beta} F_{\mu\alpha} F_{\nu \beta} - \frac{1}{2(d-1)} g_{\mu\nu} F^{\alpha \beta} F_{\alpha \beta} \right) \ .
\end{align}

We again expand the metric around the AdS metric as (\ref{expansion of metric}).
Then, the counterpart of (\ref{eq: zz}) is
\begin{align}
\Tr h'' - \frac{1}{z} \Tr h' = -16 \pi G_N \left( \frac{d-2}{d-1} z^2 \delta^{\mu\nu} F_{z\mu} F_{z\nu} - \frac{1}{2(d-1)} z^2  F_{\mu \nu} F_{\mu\nu}  \right) \ ,
  \label{eq: zz A}
\end{align}
while the counterpart of (\ref{eq: munu}) is
\begin{align}
&h_{\mu\nu}'' - (d-1) \frac{1}{z} h_{\mu\nu}' - \frac{1}{z} \Tr  h' \delta_{\mu\nu}\n
  &= -16 \pi G_N \left(  F_{z \mu} F_{z \nu} - \frac{1}{d-1} \delta_{\mu\nu} \delta^{\alpha \beta} F_{z \alpha} F_{z\beta} + \delta^{\alpha \beta} F_{\mu\alpha} F_{\nu \beta} - \frac{1}{2(d-1)} \delta_{\mu\nu} F^{\alpha \beta} F_{\alpha \beta} \right) \ .
  \label{eq: munu A}
\end{align}
The trace part of (\ref{eq: munu A}) is
\begin{align}
& \Tr h'' - (2d-1) \frac{1}{z} \Tr h'
  = -16 \pi G_N z^2\left( \frac{1}{1-d} \delta^{\mu\nu} F_{z\mu} F_{z\nu} + \frac{d-2}{2(d-1)} F^{\mu\nu} F_{\mu\nu} \right) \ ,
  \label{eq: Trmunu A}
\end{align}
and the $00$ component of  (\ref{eq: munu A}) is
\begin{align}
  & h_{00}''- (d-1) \frac{1}{z} h_{00}' - \frac{1}{z} \Tr  h'\n
    &= -16 \pi G_N \left(  F_{z 0} F_{z 0} - \frac{1}{d-1}  \delta^{\alpha \beta} F_{z \alpha} F_{z\beta} + \delta^{\alpha \beta} F_{0\alpha} F_{0 \beta} - \frac{1}{2(d-1)} F^{\alpha \beta} F_{\alpha \beta}
    \right)  \ .
  \label{eq: 00 A}
\end{align}
Taking an appropriate linear combination of (\ref{eq: zz A}), (\ref{eq: Trmunu A}) and
(\ref{eq: 00 A}) and using the equations of motion, we obtain
\begin{align}
  &\tr h'' - \frac{d - 1}{z} \tr h' \n
  &= -16 \pi G_N \left\{\frac{1}{2} \del_z (z^2 A^{\alpha} F_{z\alpha}) - \frac{d-1}{2} z A^{\alpha} F_{z\alpha} - \del_z (z^2 A_0 \del_z A_0) + (d-1)zA_0 \del_z A_0\right\} \ ,
\label{linear combination: vector}
\end{align}
where we have ignored the total derivative terms with respect to $\vec{x}$.
Integrating (\ref{linear combination: vector}) leads to
\begin{align}
  \tr h' = -8 \pi G_N z^2 A^{i}  F_{z i}
  \label{eq: trh(1)' vector}
\end{align}
where the boundary condition $\lim_{z\rightarrow\infty} h_{\mu\nu}=0$ has been used again.

 From (\ref{on-shell action for vector}) and (\ref{eq: trh(1)' vector}), we again obtain
(\ref{eq: phi on-shell}). Thus, since we have (\ref{on-shell action and information metric})
and (\ref{eq: phi on-shell}), we obtain the same formula (\ref{formula 2}) as in the case of
scalar field.

\section{Tensor field}
\setcounter{equation}{0}
Finally, let us consider the case in which a CFT is perturbed by the energy momentum tensor:
\begin{align}
  \mathcal{L} = \mathcal{L}_{CFT} + \hat{h}_{\mu\nu}(\vec{x}) T^{\mu\nu}(x) \ ,
\end{align}
where the source $\hat{h}_{\mu\nu}$ is independent of the time and $\hat{h}_{0\mu} = 0$ .
Since $T_{\mu\mu}=0$, we can assume without loss of generality that $\hat{h}_{\mu\mu}=0$.
The 2-point function of $T_{\mu\nu}$ is given by
\begin{align}
  \braket{T_{\mu\nu}(\tau,\vec{x})T_{\rho\sigma}(\tau',\vec{x}')}=\frac{C_T P_{\alpha\beta\rho\sigma}}{(\epsilon^2 + |x-x'|^2)^{d-1}} J_{\mu\alpha}(\epsilon,\tau-\tau',\vec{x}-\vec{x}') J_{\nu\beta}(\epsilon,\tau-\tau',\vec{x}-\vec{x}') \ ,
\end{align}
where
\begin{align}
&  J_{\mu\nu}(\epsilon,\tau-\tau',\vec{x}-\vec{x}')=\delta_{\mu\nu} - \frac{2(x-x')_{\mu} (x-x')_{\nu}}{\epsilon^2 + |x-x'|^2} \ , \nonumber\\
&  P_{\alpha\beta\rho\sigma}
  = \frac{1}{2} \left( \delta_{\alpha \rho} \delta_{\beta \sigma} + \delta_{\alpha \sigma} \delta_{\beta\rho} \right) - \frac{1}{d} \delta_{\alpha \beta} \delta_{\rho\sigma} \ ,
\end{align}
and $C_T$ is a normalization constant. Note that (\ref{1pt function}) and (\ref{time reversal symmetry}) are satisfied.

The information metric, which is the counterpart of (\ref{information metric 2}), is
\begin{align}
  \mathcal{G} &=  \frac{1}{2T} \int_0^{\infty} d\tau \int_{-\infty}^0  d\tau' \int d^{d-1} x \int d^{d-1} x' \hat{h}^{ij}(\vec{x}) \braket{T_{ij}(\tau,\vec{x})T_{kl}(\tau',\vec{x}')} \hat{h}^{kl}(\vec{x}')\n
  &=  \frac{C_T}{8} \int_{-\infty}^{\infty}  ds \int d^{d-1} x \int d^{d-1} x' \hat{h}^{ij}(\vec{x})\hat{h}^{kl}(\vec{x}') \frac{J_{i\alpha}(\epsilon,s,\vec{x}-\vec{x}') J_{j\beta}(\epsilon,s,\vec{x}-\vec{x}') P_{\alpha\beta kl}}{(\epsilon^2 +s^2+ (\vec{x}-\vec{x}')^2)^{d-1}} \ .
  \label{eq: information metric of tensor}
\end{align}

The bulk field corresponding to $T_{\mu\nu}$ is $h_{\mu\nu}$ in (\ref{expansion of metric}).
We solve the Einstein equation derived from $S_G$ (\ref{gravity action}) with respect to
$h_{\mu\nu}$. The boundary condition for $h_{\mu\nu}(\epsilon,x)=\hat{h}_{\mu\nu}(\vec{x})$.
We expand $h_{\mu\nu}$ as
\begin{align}
h_{\mu\nu} = h_{(1)\mu\nu}+h_{(2)\mu\nu} + \cdots \ ,
\end{align}
where $h_{(1)\mu\nu}$ and $h_{(2)\mu\nu}$ are contributions of the first and second orders
 in $\hat{h}_{\mu\nu}$, respectively.
Putting $\mathcal{T}_{MN}=0$ in
(\ref{zz component}), (\ref{zmu component}) and (\ref{munu component}).
yields the Einstein equation in this case.

By expanding the Einstein equation up to $\mathcal{O}(h_{(1)\mu\nu})$, we obtain
the equations for $h_{(1)\mu\nu}$: (\ref{zz component}) reduces to
\begin{align}
\Tr h_{(1)}''- \frac{1}{z} \Tr h'_{(1)}  = 0 \ .
  \label{eq:EoM of h(1)zz}
\end{align}
(\ref{zmu component}) reduces  to
\begin{align}
  (\del_{\mu} \Tr h_{(1)} - \del_{\nu} {h_{(1)}}^{\nu}_{\mu})' = 0  \ .
\label{eq:EoM of h(1)zmu}
\end{align}
(\ref{munu component}) reduces to
\begin{align}
h_{(1)\mu\nu}'' - (d-1) \frac{1}{z}h'_{(1)\mu\nu}- \frac{1}{z} \Tr h'_{(1)} \delta_{\mu\nu}
= \del^{\alpha} \left( \partial_{\mu} h_{(1)\nu \alpha} + \partial_{\nu}  h_{(1)\mu \alpha} - \partial_{\alpha} h_{(1)\mu \nu}\right) - \del_{\mu} \del_{\nu} \Tr h_{(1)} \ .
\label{eq:EoM of h(1)munu}
\end{align}
The solution to this equation is given by
\begin{align}
\Tr h_{(1)}'=0 \ ,  \;\;\;  \Tr h_{(1)}=0 \ .
\label{trh(1)}
\end{align}

The solution to (\ref{trh(1)}), (\ref{zmu component}) and (\ref{munu component}) with
the boundary condition $h_{(1)\mu\nu}(\epsilon,x) =\hat{h}_{\mu\nu}(\vec{x})$
is given by\cite{Liu:1998bu}
\begin{align}
  h_{(1)\mu\nu} (z, x) &=
  \alpha_T \int d^d x' \frac{z^{d}}{(z^2 + |x-x'|^2)^{d}}J_{\mu\rho}(z,\tau-\tau',\vec{x}-\vec{x}') J_{\nu\sigma}(z,\tau-\tau',\vec{x}-\vec{x}')  \nonumber\\
  & \qquad\quad \;\;\;  \times  P_{\rho\sigma ij} \hat{h}_{ij}(\vec{x}') \ .
  \label{eq: bb tensor}
\end{align}
with
\begin{align}
   \alpha_T = \frac{d+1}{d-1}\frac{\Gamma(d)}{\pi^{\frac{d}{2}} \Gamma(\frac{d}{2})}  \ .
\end{align}
Note that $h_{\mu \nu}(z, x)$ is independent of $\tau$ and that $h_{0i} = 0$.

By using (\ref{eq:EoM of h(1)zz}) and (\ref{eq:EoM of h(1)munu}), we obtain the on-shell action
for the gravitational field $h_{(1)\mu\nu}$:
\begin{align}
    &S_{G, on-shell}=\frac{1}{16\pi G_N}\int_{z=\epsilon} d^{d} x  z^{-d+1} \left\{ -\frac{1}{4}  {h_{(1)}}^{ij}{h_{(1)}}'_{ij} -\frac{1}{4}  {h_{(1)}}_{00}{h_{(1)}}'_{00} \right\}
     \label{gravity on-shell action}
\end{align}
Substituting (\ref{eq: bb tensor}) into (\ref{gravity on-shell action}) yields
\begin{align}
  S_{G, on-shell} =-\frac{d \alpha_TT}{64\pi G_N}\int^{\infty}_{-\infty} ds \int d^{d-1} x d^{d-1} x' \hat{h}_{ij}(\vec{x}) \frac{J_{i\alpha}(\epsilon,s,\vec{x}-\vec{x}')J_{j\beta}(\epsilon,s,\vec{x}-\vec{x}')P_{\alpha\beta k l}}{(\epsilon^2 +s^2+ |\vec{x}-\vec{x}'|^2)^{d-1}} \hat{h}_{kl}(\vec{x}')  \ ,
\end{align}
which implies that $C_T = \frac{d\alpha_T}{32\pi G_N}$. We find
the same relation  (\ref{on-shell action and information metric})
between the information metric and the on-shell action.


Next, we consider the equations for the back reaction $h_{(2)}$.
In what follows, we ignore total derivatives with respect to $\vec{x}$.
(\ref{zz component}) reduces to
\begin{align}
\Tr h''_{(2)} - \frac{1}{z} \Tr h'_{(2)} =
  {h_{(1)}}^{\mu\nu} {h_{(1)}}''_{\mu\nu} - \frac{1}{z} {h_{(1)}}^{\mu\nu} {h_{(1)}}'_{\mu\nu} + \frac{1}{2} {h'_{(1)}}^{\mu\nu} {h_{(1)}}'_{\mu\nu}
  \label{eq: zz g}
\end{align}
while (\ref{munu component}) reduces to
\begin{align}
& h_{(2)\mu\nu}'' - (d-1) \frac{1}{z}h'_{(2)\mu\nu}- \frac{1}{z} \Tr h'_{(2)} \delta_{\mu\nu} - 2{\rm{Ric}}^{(1)}(h_{(2)})_{\mu\nu}\n
  &
  = {h_{(1)}}'_{\mu\alpha} \delta^{\alpha\beta} {h_{(1)}}'_{\beta\nu} - \frac{1}{z} {h_{(1)}}^{\alpha\beta} {h_{(1)}}'_{\alpha\beta} \delta_{\mu\nu} - \frac{1}{2}\Tr h'_{(1)} h'_{\mu\nu} + \frac{1}{z}\Tr h' h_{\mu\nu} + 2 {\rm{Ric}}^{(2)}(h_{(1)})_{\mu\nu}
\label{munu component of equation for h2}
\end{align}
The trace part and the 00 component of
(\ref{munu component of equation for h2}) takes the forms
\begin{align}
\Tr h_{(2)}'' - (2d-1) \frac{1}{z} \Tr h_{(2)}'
  & = \frac{1}{2}{h_{(1)}}^{\mu\nu} {h_{(1)}}''_{\mu\nu} + {h_{(1)}}'^{\mu\nu} {h_{(1)}}'_{\mu\nu} - \frac{3d - 1}{2z} {h_{(1)}}^{\mu\nu} {h_{(1)}}'_{\mu\nu}  \ ,
  \label{eq: Trmunu g}
\end{align}
\begin{align}
h_{(2) 00}''- (d-1) \frac{1}{z} h_{(2) 00}' - \frac{1}{z} \Tr  h_{(2)}'
  =& {h_{(1)}}_{0}^{\alpha} {h''_{(1)}}_{0\alpha} + {h'_{(1)}}_{0}^{\alpha} {h'_{(1)}}_{0\alpha} -\frac{d-1}{z}{h_{(1)}}_{0}^{\alpha} {h'_{(1)}}_{0\alpha} \n
  &- \frac{1}{z}{h_{(1)}}^{\mu\nu} {h_{(1)}}_{\mu\nu} \ ,
  \label{eq: 00 g}
\end{align}
respectively, and
(\ref{eq:EoM of h(1)zz}), (\ref{eq:EoM of h(1)zmu}),  (\ref{eq:EoM of h(1)munu}) and
(\ref{trh(1)}) have been used.
Taking appropriate linear combinations of (\ref{eq: zz g}), (\ref{eq: Trmunu g}) and
(\ref{eq: 00 g}), we obtain an equation for $\tr  h'_{(2)} = \delta^{ij} h'_{(2)ij}$:
\begin{align}
  \tr  h'_{(2)} = \frac{3}{4} {h_{(1)}}^{i j}{h'_{(1)}}_{i j}
  - \frac{1}{4} {h_{(1)}}_{00} {h'_{(1)}}_{00} \ .
  \label{sol: h(2)}
\end{align}

By using (\ref{sol: h(2)}), the gravity on-shell action (\ref{gravity on-shell action}) is rewritten as
\begin{align}
  S_{G, on-shell}=& \frac{1}{16\pi G_N}\int_{z=\epsilon} d^{d} x z^{-d+1}\left\{ \tr h'_{(2)} -  {h_{(1)}}^{ij} {h_{(1)}}'_{ij}  \right\} \ .
  \label{eq: g on-shell}
\end{align}
The first term comes from the $S_{EH}$, being a counterpart of
(\ref{eq: phi on-shell}).

The induced metric of a hypersurface with $\tau$ and $z$ fixed is
given by $\gamma_{ij} = z^{-2} (\delta_{ij} + h_{(1)ij}+h_{(2)ij})$
The volume of the hypersurface is
\begin{align}
  v &= \int d^{d-1} x \sqrt{\gamma}\n
  &= \int d^{d-1} x z^{-d+1} (1 - \frac{1}{4}{h_{(1)}}^{ij} {h_{(1)}}_{ij} + \frac{1}{2} \tr h_{(2)})
\end{align}
The counterpart of (\ref{derivative of delta v}) is given by
\begin{align}
  \delta v' &= \frac{1}{2}\int d^{d-1} x \left\{ (-d+1)z^{-d} (\tr h_{(2)} - \frac{1}{2}{h_{(1)}}^{ij} {h_{(1)}}_{ij}) + z^{-d+1}( \tr h'_{(2)} - {h_{(1)}}^{ij} {h_{(1)}}'_{ij}) \right\}
\label{derivative of delta v: gravity}
\end{align}
While the first and second terms in (\ref{derivative of delta v: gravity}) represent the canonical scaling of the volume,
the third and fourth terms represent a nontrivial scaling of the volume and is proportional to
the righthand side of (\ref{eq: g on-shell}).
Thus, from (\ref{eq: g on-shell}), we again obtain (\ref{formula 2}) with
(\ref{delta v nontrivial}) replaced by
\begin{align}
\delta v'_{nontrivial} = \frac{1}{2}\int_{z=\epsilon}d^{d-1} x \
   z^{-d+1}( \tr h'_{(2)} - {h_{(1)}}^{ij} {h_{(1)}}'_{ij})  \ .
\end{align}


\section{Conclusion and discussion}
\setcounter{equation}{0}

In this paper, we studied how information geometry is described by bulk geometry.
We considered a quantum information metric that measures the distance between
the ground states of a CFT and a theory obtained by perturbating the CFT.
We represented the information metric in terms of
the back reaction that the bulk geometry gains
due to the perturbation.
We found the formula (\ref{formula 2}) that expresses the information
metric by deviation of the volume of the hypersurface in the bulk from that in the AdS case.
The geometrical quantity is local in the bulk direction.
This formula is universal in the sense that it holds for all the cases
of scalar, vector and tensor perturbations. It associates information geometry with
dynamics of gravity.
The information metric is related  to a codimension-2 hypersurface in this paper, while it is related to
a codimension-1 hypersurface in \cite{MIyaji:2015mia,Bak:2015jxd,Trivella:2016brw,Chen:2018vkw,
Karar:2019wjb}. This difference comes from one between the situations considered on the gravity side
as mentioned in section 1.

We associated the information metric with the volume of a hypersurface specified by
$z=\epsilon$ with $\epsilon$ small and $\tau=$ const..
In order to reconstruct full bulk geometry from field theory,
we should associate it with a hypersurface specified by $z=$ an arbitrary constant
by using renormalization group.
Furthermore, it is needed to find
relationship between information geometry and
bulk quantities local even in $\vec{x}$ directions.
To understand the geometrical meaning of the formula (\ref{formula 2})  more deeply, we should find how  general codimension-2 hypersurfaces in bulk are related to  information geometry.
It is also relevant to derive effects of strings and quantum gravity from information
geometry to construct quantum theory of gravity.
We hope to report progress in these issues in the near future.


\section*{Acknowledgements}
A.T.\ was supported in part by Grant-in-Aid for Scientific Research
(No. 18K03614) from Japan Society for the Promotion of Science.

\appendix

\renewcommand{\theequation}{A.\arabic{equation}}
\setcounter{equation}{0}

\section{Ricci tensor and scalar curvature}
\label{sec:appendix A}
In this appendix, we calculate the Ricci tensor and the scalar curvature for the metric (\ref{metric}).

The Christoffel symbols are given by
\begin{align}
\Gamma^{z}_{zz}&=-\frac {1}{z^{2}}   \ , \\
\Gamma ^{z}_{\alpha \beta }&  =\frac {1}{z}g_{\alpha \beta }-\frac{1}{2}g'_{\alpha \beta} \ , \\
\Gamma ^{\alpha}_{z \gamma }& =\frac {1}{2}g^{\alpha \beta }g'_{\beta \gamma }-\frac {1}{z}\delta ^{\alpha }_{\gamma } \ , \\
\Gamma ^{\alpha}_{\beta \gamma }& = \frac {1}{2}g^{\alpha \delta}\left( \partial _{\beta } g_{\gamma \delta}+\partial _{\gamma}g_{\beta\delta }-\partial _{\delta} g_{\beta \gamma  }\right)\n
&= \Gamma ^{\alpha}_{\beta \gamma }(g) \ , \\
\Gamma ^{z}_{\alpha z } &= \Gamma ^{\alpha}_{zz} = 0 \ .
\end{align}

The Riemann curvature, the Ricci tensor and the scalar curvature are defined by
\begin{align}
&R^I_{\;\;JKL}[G] = \del_{K} \Gamma^I_{JL} - \del_{L} \Gamma^I_{JK} + \Gamma^I_{MK} \Gamma^M_{LJ} - \Gamma^I_{ML} \Gamma^M_{KJ}    \ ,  \label{Riemann curvature} \\
& R_{MN}[G] = R^K_{\;\; MKN}[G] \ ,  \label{Ricci tensor}  \\
& R[G] = G^{MN} R_{MN}[G]  \ ,  \label{scalar curvature}
\end{align}
respectively.

The components of the Ricci tensor are given by
\begin{align}
  R[G]_{zz}
  &= -\frac{1}{2}\left(  \Tr g^{-1} g'' - \frac{1}{z} \Tr g^{-1} g' - \frac{1}{2} \Tr g^{-1} g'g^{-1} g' + \frac{2}{z^2}  \right)\ , \\
  R[G]_{z\mu}
   &= \frac{1}{2} \left( \nabla^{\alpha} g'_{\alpha \mu} - \nabla_{\mu} \Tr g^{-1} g' \right) \ , \\
  R[G]_{\mu \nu}
  &=  - \frac{1}{2} \left( - 2 {\rm{Ric}}_{\mu\nu}(g) + g''_{\mu\nu} - g'_{\mu \lambda} g^{\lambda \sigma} g'_{\sigma \nu} + \frac{1}{2} \Tr \left(g^{-1} g' \right) g'_{\mu\nu} \right. \n
  &\ \ \ \ \ \ \ \ \ \ \ \ \ \ \ \ \ \ \ \ \ \ \ \ \ \ \ \ \ \ \ \ \ \left. - (d-1) \frac{1}{z} g'_{\mu\nu} - \frac{1}{z} \Tr \left( g^{-1} g' \right) g_{\mu\nu} + \frac{2d}{z^2} g_{\mu\nu} \right)
\ ,
\end{align}
where
\begin{align}
{\rm{Ric}}(g)_{\mu \nu} &= \del_{\alpha} \Gamma^{\alpha}_{\mu \nu}(g)
- \del_{\nu} \Gamma^{\alpha}_{\mu \alpha}(g)
+ \Gamma^{\alpha}_{\beta \alpha}(g) \Gamma^{\beta}_{\nu \mu}(g)
- \Gamma^{\alpha}_{\mu \beta}(g) \Gamma^{\beta}_{\nu \alpha}(g) \ .
\end{align}
The scalar curvature is given by
\begin{align}
  R[G] &= G^{MN} R[G]_{MN} = G^{zz} R[G]_{zz} + G^{\mu\nu} R[G]_{\mu\nu}\n
    &= z^2 \left( - \Tr g^{-1} g'' + \frac{3}{4} \Tr g^{-1} g'g^{-1} g' + \frac{d}{z} \Tr g^{-1} g' - \frac{1}{4} \left( \Tr g^{-1} g' \right)^2 - \frac{d^2 + d}{z^2} + R(g)\right) \ ,
\end{align}
where $R(g)=g^{\mu\nu}{\rm{Ric}}(g)_{\mu \nu}$.

We expand ${\rm{Ric}}(g)_{\mu\nu}$ with respect to $h_{\mu\nu}$ in (\ref{expansion of metric}):
\begin{align}
  {\rm{Ric}}(g)_{\mu \nu} = {\rm{Ric}}^{(1)}(h)_{\mu \nu} + {\rm{Ric}}^{(2)}(h)_{\mu \nu} + ...
\end{align}
where the first order of $h_{\mu\nu}$, ${\rm{Ric}}^{(1)}(h)_{\mu \nu}$, is
\begin{align}
{\rm{Ric}}^{(1)}(h)_{\mu \nu}
& = \frac{1}{2} \del^{\alpha} \left( \partial_{\mu} h_{\nu \alpha} + \partial_{\nu}  h_{\mu \alpha} - \partial_{\alpha} h_{\mu \nu}\right) - \frac{1}{2} \del_{\mu} \del_{\nu} h_{ \alpha}^{\alpha}
\end{align}
and the second order of $h_{\mu\nu}$, $ {\rm{Ric}}^{(2)}(h)_{\mu\nu}$, is
\begin{align}
  {\rm{Ric}}^{(2)}(h)_{\mu\nu}  
  =&
   -  \frac{1}{2} \del_{\alpha} \left\{ h^{\alpha \beta} \left( \partial_{\mu} h_{\nu \beta} + \partial_{\nu}  h_{\mu \beta} - \partial_{\beta} h_{\mu \nu}\right) \right\} \n
  &
    + \frac{1}{2} \del_{\nu} \left\{ h^{\alpha \beta}  \partial_{\mu} h_{\alpha \beta}\right\} + \frac{1}{4}  \partial^{\beta} h_{\alpha }^{\alpha} \left( \partial_{\mu} h_{\nu \beta} + \partial_{\nu}  h_{\mu \beta} - \partial_{\beta} h_{\mu \nu}\right)\n
  &- \frac{1}{4} \delta^{\alpha \gamma} \delta^{\beta \delta}
  \left( \partial_{\mu} h_{\beta \gamma} \partial_{\mu} h_{\alpha \delta} + \partial_{\beta}  h_{\mu \gamma}\partial_{\alpha}  h_{\mu \delta} \right.\n
  &\left. \ \ \ \ \ \ \ \ \ \ \ \ \ \ \ \
   - \partial_{\beta}  h_{\mu \gamma} \partial_{\delta} h_{\mu \alpha} - \partial_{\gamma} h_{\mu \beta} \partial_{\delta} h_{\mu \alpha} + \partial_{\gamma} h_{\mu \beta} \partial_{\delta} h_{\mu \alpha}\right).
\end{align}
Then, we obtain  the second order of $h_{\mu\nu}$ in $R(g)$,
\begin{align}
  R^{(2)}(h) =& \delta^{\mu\nu} {\rm{Ric}}^{(2)}(h)_{\mu\nu} - {h}^{\mu\nu} {\rm{Ric}}^{(1)}(h)_{\mu\nu} \n
  =&
  -  \frac{1}{2} \del_{\alpha} \left\{ {h}^{\alpha \beta} \left( 2 \partial^{\mu} {h}_{\mu \beta} - \partial_{\beta} {h}_{\mu}^{\mu}\right) \right\} \n
  &
    + \frac{1}{2} \del^{\mu} \left\{ {h}^{\alpha \beta}  \partial_{\mu} {h}_{\alpha \beta} \right\} + \frac{1}{4}  \partial^{\beta}  {h}_{\alpha }^{\alpha} \left( 2 \partial^{\mu} {h}_{\mu \beta} - \partial_{\beta} {h}_{\mu}^{\mu}\right)\n
  &- \frac{1}{4}
  \left( 2 \partial^{\alpha}  {h}^{\mu \nu} \partial_{\nu}  {h}_{\mu \alpha}  - \partial_{\alpha} {h}^{\mu \nu} \partial_{\alpha} {h}_{\mu \nu} \right) \n
  & - {h}^{\mu\nu} \left\{ \frac{1}{2} \del^{\alpha} \left( 2  \partial_{\mu} h_{\nu \alpha} - \partial_{\alpha} h_{\mu \nu}\right) - \frac{1}{2} \del_{\mu} \del_{\nu} {h}_{\alpha}^{\alpha} \right\}.
  \label{def: second order of scalar curvature}
\end{align}

We expand the Einstein-Hilbert action and the Gibbons-Hawking term up tp ${\mathcal{O}(h^2)}$:
\begin{align}
  S_{EH}
  = & \frac{1}{16\pi G_N} \int d^{d+1} x \sqrt{G} \left(- R[G] + 2\Lambda \right) \n
  = & \frac{1}{16\pi G_N}\int d^{d+1} x z^{-d+1}  \left\{ - \frac{2d}{z^2} +  {h''}^{\mu}_{\mu} - \frac{d}{z} {h'}^{\mu}_{\mu} + \frac{d}{z^2} {h}^{\mu}_{\mu} \right.\n
  & - h^{\mu\nu} {h''} _{\mu\nu}- \frac{3}{4}   {h'}^{\mu\nu}{h'}_{\mu\nu} +\frac{d}{z}{h}^{\mu\nu} {h'}_{\mu\nu} - \frac{d}{2z^2}{h}^{\mu\nu} {h}_{\mu\nu} + \frac{1}{4} \del^{\alpha} h^{\mu\nu} \del_{\alpha} h_{\mu\nu}- \frac{1}{2} \del^{\alpha} h^{\mu\nu} \del_{\mu} h_{\mu\alpha} \n
  &\left. + \frac{1}{2} h_{\mu}^{\mu} {h''}_{\nu}^{\nu} + \frac{1}{4} ({h'}_{\mu}^{\mu})^2 - \frac{d}{2z} h_{\mu}^{\nu} {h'}_{\nu}^{\nu} + \frac{d}{4z^2} ({h}_{\mu}^{\mu})^2 + \frac{1}{2} \del_{\nu} h^{\mu\nu} \del_{\mu} {h}_{\alpha}^{\alpha} - \frac{1}{4} \del^{\alpha} {h}_{\mu}^{\mu} \del_{\alpha} {h}_{\nu}^{\nu} \right\} \\
  S_{GH}
  = & -\frac{1}{16\pi G_N} \int_{z=\epsilon} d^d x \sqrt{\gamma} (2K+\lambda)  \n
  =& \frac{1}{16\pi G_N}\int_{z=\epsilon} d^d x z^{-d} \left\{ -2 - h^{\mu}_{\mu} - \frac{1}{4} (h^{\mu}_{\mu})^2 + \frac{1}{2} h^{\mu\nu} h_{\mu\nu} - h^{\mu\nu}{h'}_{\mu\nu} + {h'}^{\mu}_{\mu} +\frac{1}{2} h^{\mu}_{\mu} {h'}^{\nu}_{\nu} \right\} \ .
\end{align}

By using the Einstein equations to the first order of $h_{\mu\nu}$, (\ref{eq:EoM of h(1)zz}),
(\ref{eq:EoM of h(1)zmu}), (\ref{eq:EoM of h(1)munu}) and (\ref{trh(1)}),  the Einstein-Hilbert action $S_{EH}$ is reduced to
\begin{align}
  S_{EH, on-shell}
  =\frac{1}{16\pi G_N}\int d^{d+1} x \frac{d}{dz} \left\{- \frac{3}{4} z^{-d+1}  {h}^{\mu\nu}{h}'_{\mu\nu} + \frac{1}{2}z^{-d} {h}^{\mu\nu} {h}_{\mu\nu} - 2 z^{-d} \right\}
  \label{EH on-shell action}
  \end{align}
and the Gibbons-Hawking term $S_{GH}$ is reduced to
\begin{align}
    &S_{GH, on-shell}=
     \frac{1}{16\pi G_N}\int_{z=\epsilon} d^d x z^{-d} \left\{ - z {h}^{\mu\nu}  h'_{\mu\nu} + \frac{1}{2} {h_{(1)}}^{\mu\nu}{h_{(1)}}_{\mu\nu} -2 \right\}
    %
    \label{GH on-shell action}
\end{align}
Thus, using (\ref{EH on-shell action}) and (\ref{GH on-shell action}), we obtain
the on-shell action (\ref{gravity on-shell action}) on the boundary specified by $z=\epsilon$.

\end{document}